\documentstyle[psfig,pre,multicol,aps]{revtex}

\def\sech{\mbox{sech}}

\begin{document}
\draft

\title{Kinks in the discrete sine-Gordon model \\
with Kac-Baker long-range interactions}
\author{Serge F. Mingaleev $^{a, b}$ and Yuri B. Gaididei}
\address{$^{a}$ Bogolyubov Institute for Theoretical Physics, 
03143 Kiev, Ukraine}
\author{Eva Majern{\'\i}kov{\'a}}
\address{$^{b}$ Department of Theoretical Physics of Palack{\'y} 
University, CZ-77207 Olomouc, Czech Republic}
\address{$^{c}$ Institute of Physics, SAS, SK-84228 Bratislava, 
Slovak Republic}
\author{Serge Shpyrko $^{d, b}$}
\address{$^{d}$ Institute for Nuclear Research, 252028 Kiev, Ukraine}

\date{\today}

\maketitle

\begin{abstract}
We study effects of Kac-Baker long-range dispersive interaction 
(LRI) between particles on kink properties in the discrete 
sine-Gordon model. We show that the kink width increases 
indefinitely as the range of LRI grows only in the case 
of strong interparticle coupling. 
On the contrary, the kink becomes {\em intrinsically 
localized} if the coupling is under some critical value. 
Correspondingly, the Peierls-Nabarro barrier vanishes as 
the range of LRI increases for supercritical values of 
the coupling but remains {\em finite} for subcritical values. 
We demonstrate that LRI essentially transforms the internal 
dynamics of the kinks, specifically creating their 
{\em internal localized} and {\em quasilocalized modes}. 
We also show that {\em moving kinks radiate} plane waves 
due to break of the Lorentz invariance by LRI.
\end{abstract}

\begin{multicols}{2}
\narrowtext

\section{Introduction}
\label{sec:intro}

The effects of long-range dispersive interactions (LRI's) on the
dynamics and thermodynamics of soliton-bearing systems have
attracted a great deal of interest in the past decade 
\cite{Ishimori:1982:PTP,Remoissenet:1985:JPC,Tchawoua:1993:JPA,Neuper:1994:PLA,Gaididei:1995:PRL,Gaididei:1997:PD,Bonart:1997:PLA,Flach:1998:PRE,Mingaleev:1998:PRE,Gaididei:1996:PLA,Gaididei:1996:PS,Gaididei:1997:PRE,Rasmussen:1998:PD,Johansson:1998:PRE,Cruzeiro:1998:PLA,Braun:1990:PRB,Sarker:1981:PRB,Woafo:1991:JPCM,Woafo:1992:JPCM,Woafo:1993:PRB,Woafo:1993:JPCM,Kenne:1994:JPCM}. 
Such attention is due to the fact that in realistic physical
systems the interparticle forces are always long-ranged to some 
extent, and if the range of LRI's exceeds some critical value, 
they change soliton features {\em qualitatively}. 
In particular, the competition between short-range and long-range 
interactions in anharmonic chains 
\cite{Ishimori:1982:PTP,Remoissenet:1985:JPC,Tchawoua:1993:JPA,Neuper:1994:PLA,Gaididei:1995:PRL,Gaididei:1997:PD,Bonart:1997:PLA,Flach:1998:PRE,Mingaleev:1998:PRE}
and nonlinear Schr{\"o}dinger (NLS) models 
\cite{Gaididei:1996:PLA,Gaididei:1996:PS,Gaididei:1997:PRE,Rasmussen:1998:PD,Johansson:1998:PRE} 
brings into existence several types of the soliton states. In the
nonlocal discrete NLS model two types of stable soliton states can
coexist at the same excitation number \cite{Gaididei:1997:PRE}. 
In other words, there occurs a bistability phenomenon with a 
possibility of controlled switching between states 
\cite{Johansson:1998:PRE}. 
Besides, the power law LRI manifests itself in algebraic 
soliton tails 
\cite{Gaididei:1997:PRE,Flach:1998:PRE,Mingaleev:1998:PRE} 
and can give rise to an energy gap between the spectra of plane 
waves and the soliton states \cite{Mingaleev:1998:PRE}. 

In the present paper we consider the effects of LRI's in discrete 
Klein-Gordon (KG) models. 
These models were successfully used in investigations of a number 
of physical phenomena such as dislocations in solids, charge-density 
waves, adsorbed layers of atoms, domain walls in ferromagnets and 
ferroelectrics, crowdions in metals, and hydrogen-bonded molecules 
(see the review paper \cite{Braun:1998:PR} for references). 
As it is known \cite{Braun:1990:PRB,Braun:1998:PR} the interparticle 
interactions in many of these systems are 
substantially long-ranged. 

In the assumption of the harmonic interaction between particles 
the dimensionless Hamiltonian of the discrete KG model can be written 
in the form 
\begin{eqnarray}
\label{sys:hamil}
H &=& \sum_n \biggl\{ \frac{1}{2} \Bigl( \frac{du_n}{dt} \Bigr)^2 + 
V(u_n) \nonumber \\ 
&+& \frac{1}{2} \sum_{m>n} J_{m,n} (u_m-u_n)^2 \biggr\} \; , 
\end{eqnarray}
where $u_n$ is the displacement of the $n$-th particle from its
equilibrium position and $J_{m,n}$ is the coupling constant 
between particles $n$ and $m$. 

As far as we know, until now there was only one investigation 
\cite{Braun:1990:PRB} of the KG model with the power law LRI. 
It was shown that the asymptotics of the kink shape as well as 
the interaction energy of the kinks are power law and, because 
of this, the dependence of the Peierls-Nabarro 
barrier versus the atom concentration is similar to the 
``devil's staircase''. 

But KG models with the exponential law 
(usually called Kac-Baker) LRI 
\begin{equation}
\label{sys:Jmn}
J_{m,n}=J (e^{\alpha}-1) e^{-\alpha |m-n|} \; 
\end{equation}
were believed to have been investigated in an exhaustive fashion 
\cite{Sarker:1981:PRB,Woafo:1991:JPCM,Woafo:1992:JPCM,Woafo:1993:PRB,Woafo:1993:JPCM,Kenne:1994:JPCM}. 
As early as 1981, Sarker and Krumhansl found \cite{Sarker:1981:PRB} 
an analytical kink solution for the $\phi^4$ model. The width and the
energy of the kinks were found to increase indefinitely as 
$\alpha$ decreases. 
An important role of the Kac-Baker LRI in thermodynamics of the system 
was also shown. Within the decade Woafo {\em et
al.} considered in a series of papers 
\cite{Woafo:1991:JPCM,Woafo:1992:JPCM,Woafo:1993:PRB} 
the discreteness effects in the same model. They have shown that
the Peierls-Nabarro barrier vanishes as $\alpha \to 0$. 

More recently the sine-Gordon (SG) model 
\begin{equation} 
\label{sys:pot} 
V(u_n) = 1-\cos u_n \; 
\end{equation}
with Kac-Baker LRI (\ref{sys:Jmn}) has been
studied \cite{Woafo:1993:JPCM,Kenne:1994:JPCM} and all results of
Ref. \cite{Sarker:1981:PRB} have been extended to this model. 
An implicit form for the kinks has been obtained and the kink energy 
and width have been found to grow to infinity as $\alpha \to 0$ in Ref. 
\cite{Woafo:1993:JPCM}. The thermodynamics of the system has been
thereafter studied in Ref. \cite{Kenne:1994:JPCM}. 

Thus, the investigations performed in Refs. 
\cite{Sarker:1981:PRB,Woafo:1991:JPCM,Woafo:1992:JPCM,Woafo:1993:PRB,Woafo:1993:JPCM,Kenne:1994:JPCM} 
give the impression that the Kac-Baker LRI always results, in the
limit $\alpha \to 0$, into infinite increasing of the kink width (and,
therefore, vanishing of the Peierls-Nabarro barrier). However, closer
inspection shows that this conclusion is proper for the case $J>0.5$
only. 

In the present paper we explore the effects of the Kac-Baker LRI 
in the discrete sine-Gordon model (\ref{sys:hamil})-(\ref{sys:pot}) further 
so that we could cover the case $J<0.5$. 
What is more, we investigate the internal kink dynamics and the 
radiation of moving kinks. 

The paper is organized as follows. In Sec. \ref{sec:eqs} we derive the
equation of motion of the system in the continuum limit using the 
technique of pseudo-differential operators. Then, in 
Sec. \ref{sec:kink} we solve this equation and obtain an implicit
analytical form of the kink solution for arbitrary values of $\alpha$ 
and $J$. Turning back to the discrete case we 
calculate the form of the kinks numerically and compare it with
the analytical solution. We show that in the case of 
$J(e^{\alpha}+1)<1$ the kinks are {\em intrinsically localized}.
The calculation of the Peierls-Nabarro barrier 
as a function of $\alpha$ and $J$ finishes the section. 
It turns out that the Peierls-Nabarro barrier vanishes in the limit 
$\alpha \to 0$ for $J>0.5$ but {\em remains finite} for $J<0.5$. 
In Sec. \ref{sec:internal} we develop a variational approach to 
the internal kink dynamics and demonstrate that LRI strongly 
enhances creation of kink's internal modes. Then we validate this 
result by direct numerical calculations. 
We show that similar to the non-sinusoidal Peyrard-Remoissenet 
potential \cite{Peyrard:1982:PRB,Braun:1997:PRE}, the Kac-Baker
LRI (\ref{sys:Jmn}) with small $\alpha$ creates {\em several} kink's
internal modes. By this means our results support the recent 
conclusion of Kivshar {\em et al.} \cite{Kivshar:1998:PRL} that 
``the internal mode is a {\em fundamental concept} for many
nonintegrable soliton models''. 
Moreover, we show that for large values of $J$, for which kink's 
internal modes do not exist, the Kac-Baker LRI gives rise to 
pronounced {\em quasilocalized modes} inside of the phonon 
spectrum. 
In Sec. \ref{sec:radiation} we show that due to break of the 
Lorentz invariance by LRI, there are no stationary moving kinks 
in the system; arbitrary moving kink will radiate plane waves 
with the wavelength proportional to its velocity. 
In Sec. \ref{sec:summary} we summarize and discuss the obtained 
results.

\section{Equations of motion}
\label{sec:eqs}

The Hamiltonian (\ref{sys:hamil})-(\ref{sys:pot}) generates the
equation of motion 
\begin{equation}
\label{sys:eq-un}
\frac{d^2 u_n}{dt^2} - \sum_{m \neq n} J_{m,n} (u_m -u_n) + \sin u_n =
0 \; . 
\end{equation}
To obtain its solution analytically we pass to the continuum limit 
treating $n$ as a continuous variable $n \to x=an$, where $a$ is the
distance between particles. Thus, using 
\begin{equation}
\label{sys:cont}
u_n(t) \to u(x,t) \quad , \quad 
u_m(t) \to e^{(am-x) \partial_x} u(x,t) \; 
\end{equation}
and keeping formally all terms in the Taylor expansion of 
$e^{(am-x) \partial_x}$, we can cast Eq. (\ref{sys:eq-un}) in the 
operator form 
\begin{equation}
\label{sys:eq-ux}
\partial^2_t u - 
\frac{J (e^{\alpha}+1) \sinh^2 (a \partial_x /2)}{\sinh^2(\alpha/2)
- \sinh^2 (a \partial_x /2)} u + \sin u=0 \; ,
\end{equation}
where $\partial_x$ and $\partial_t$ are the derivatives
with respect to $x$ and $t$, respectively, and the identity 
\begin{equation}
\label{sys:exp-sum}
\sum_{m \neq 0} e^{-\alpha |m|+am \partial_x} \equiv \frac{\cosh(a
\partial_x) - e^{-\alpha}}{\cosh (\alpha) - \cosh(a \partial_x)} \; 
\end{equation}
has been used. 

In the approximation $\sinh(a \partial_x /2) \approx a \partial_x /2$ 
the equation of motion (\ref{sys:eq-ux}) takes on the form 
\begin{equation}
\label{sys:eq-uz}
\partial^2_t u - 
\frac{\partial_z^2}{1 - \sigma^2 \partial_z^2} u + \sin u=0 \; ,
\end{equation}
where $z=x/(a \xi)$ with 
\begin{equation}
\label{sys:xi-sigma}
\xi = \frac{\sqrt{J(e^{\alpha}+1)}}{2 \sinh(\alpha/2)} \qquad
\mbox{and} \qquad \sigma^2=\frac{1}{J(e^{\alpha}+1)} \; . 
\end{equation}
Here the parameter $\xi$ presents a measure for the soliton width --- the
continuum approximation should be good for large enough values of $\xi$. 

Acting on Eq. (\ref{sys:eq-uz}) by the operator 
$(1 - \sigma^2 \partial_z^2)$ one can write the equation of motion in
the differential form 
\begin{equation}
\label{sys:eq-uz-diff}
u_{tt} - u_{zz} + \sin u = \sigma^2 u_{zztt} + \sigma^2 (\sin u)_{zz} \; , 
\end{equation}
which coincides with the equation derived in Ref.
\cite{Woafo:1993:JPCM}. 
The authors of that paper found the form of the moving kink solution 
neglecting the term $u_{zztt}$. Thus they have found in fact an 
exact solution for immobile kinks
but approximate for moving ones. In Sec. \ref{sec:radiation} we will 
show that the term $u_{zztt}$ is responsible for the non-perturbative 
radiation of the kink. Indeed, due to break of the Lorentz invariance 
by LRI there are no stationary moving kinks in the system --- moving 
kinks always radiate and eventually stop. It is why in the next 
section we consider the immobile kinks only and write their exact 
shape in to some extent more simple and general form than that given  
in Ref. \cite{Woafo:1993:JPCM}.

\section{Kink's static properties} 
\label{sec:kink}

\subsection{Analytical kink solution} 
\label{sec:anal}

In this section we obtain the immobile ($\partial^2_t u =0$) 
kink solution of Eq. (\ref{sys:eq-uz}). 
Denoting this solution as $\phi(y;B)$, where 
\begin{eqnarray}
\label{sol:y}
y=\frac{z}{\sigma} = \frac{2}{a} \sinh \biggl( \frac{\alpha}{2} \biggr)
x \; 
\end{eqnarray}
and 
\begin{eqnarray}
\label{sol:B}
B^2= 1+ \frac{1}{\sigma^2} = 1+J(e^{\alpha}+1) \; , 
\end{eqnarray}
one can see that Eq. (\ref{sys:eq-uz}) takes on the form 
\begin{equation}
\label{sol:eq-B}
\frac{\partial_y^2}{1-\partial_y^2} \phi(y;B) 
= \frac{\sin \phi}{B^2-1} \; . 
\end{equation}
Using a new variable $v=\sin(\phi/2)$ one can rewrite it in the 
polynomial form 
\begin{eqnarray}
\label{sol:eq-v}
(1-v^2)(B^2-2 v^2) v_{yy} \nonumber \\
+ v(B^2-4+2v^2) v_y^2 = v(1-v^2)^2 \; , 
\end{eqnarray}
and multiplying by $(B^2-2v^2)/(1-v^2)^2$ one can cast it in the 
form of the equation of motion of certain Hamiltonian system 
with the Hamiltonian 
\begin{equation}
\label{sol:hamil-v}
h= \frac{(B^2-2v^2)^2}{2(1-v^2)} v_y^2 - \frac{1}{2} v^2 (B^2-v^2) \; . 
\end{equation}
Imposing the kink's boundary conditions ($v_y \to 0$ when $v \to 0$)
we arrive at the constraint $h=0$. Thus we obtain the equation 
\begin{equation}
\label{sol:der-v}
v_y^2 = \frac{v^2 (1-v^2) (B^2-v^2)}{(B^2-2v^2)^2} \; , 
\end{equation}
which after integration gives the kink solution of the form 
\begin{eqnarray}
\label{sol:sol-y}
\pm (y-y_0) &=& \frac{B}{2} \log \biggl( \frac{1+B\mu}{1-B\mu} \biggr) 
+ \log \biggl( \frac{1-\mu}{1+\mu} \biggr) \nonumber \\ 
&=& \sum_{m=1}^{\infty} \frac{B^{2m} -2}{2m-1} \mu^{2m-1} 
\; , 
\end{eqnarray}
where a new variable 
\begin{equation}
\label{sol:mu}
\mu = \sqrt{\frac{1-v^2}{B^2-v^2}} \;  
\end{equation}
is introduced. Turning back to the function $\phi(y;B)$ we 
obtain an exact form of the kink (positive
sign in Eq. (\ref{sol:sol-y})) or antikink 
(negative sign) centered at  $y_0$:
\begin{equation}
\label{sol:sol-phi}
\phi(y;B)=2 \arcsin \biggl( \sqrt{\frac{1-B^2 \mu^2}{1-\mu^2}} \biggr) \; , 
\end{equation}
where the dependence of $\mu$ on $y$ is determined by Eq. 
(\ref{sol:sol-y}). 

It can be checked that in the nearest-neighbors interaction (NNI) 
limit ($B \to \infty$) the solution reduces to the ordinary SG 
kink or antikink form
\begin{equation}
\label{sol:sol-u-local}
u(z)=4 \arctan(\exp(\pm (z-z_0))) \; , 
\end{equation}
where we have used the variables $u(z)$ and $z$ defined in the
previous section. 

\begin{figure}
\centerline{\hbox{
\psfig{figure=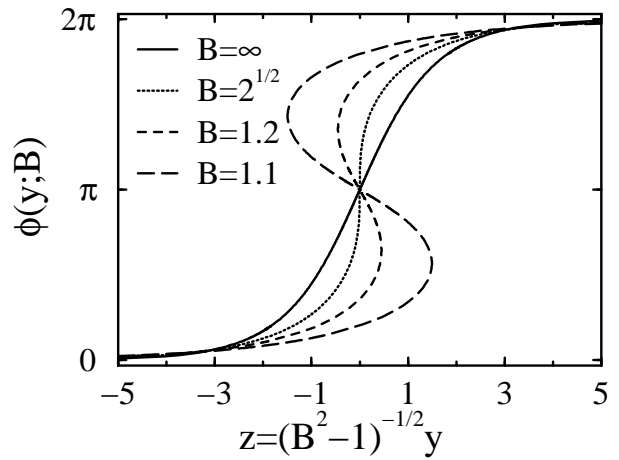,width=80mm,angle=0}}}
\caption{
The kink shape predicted analytically for different values 
of the range of LRI: it is usual in the NNI limit 
($B=\infty$, full line), has the vertical slope in the 
critical case ($J(e^{\alpha}+1)=1$ or $B=\sqrt{2}$, 
dotted line), and is multivalued (${\cal S}$-shaped) in the 
supercritical cases of $B=1.2$ (dashed line) and 
$B=1.1$ (long-dashed line).} 
\label{fig:phi-y-form}
\end{figure}

Looking at Eq. (\ref{sol:sol-y}) one can see that near the center of
the kink, where $\mu$ is small, the first order term in the
Taylor series vanishes at $B=\sqrt{2}$. In this case the derivative 
$d \mu / dy = \pm 1 / 2\mu^2$ 
goes to infinity in the center ($\mu \to 0$) of the kink. It means
that the slope of the kink becomes vertical for $B=\sqrt{2}$ (see
Fig. \ref{fig:phi-y-form}). If $B<\sqrt{2}$ the slope of the kink
assumes negative values and the solution (\ref{sol:sol-phi}) 
becomes ${\cal S}$-shaped (multivalued) and thus loses its 
physical meaning. 

Thus, returning to the initial physical parameters $J$ and $\alpha$ we
can state that there exists a critical value of $J$ in the system: if 
$J>0.5$ the value
of $B$ always exceeds $\sqrt{2}$ and the form of the kink does not
change drastically with $\alpha$ (see Fig. \ref{fig:u-n-form}-a). 
It was this case which was 
studied in details in Refs. \cite{Woafo:1993:JPCM,Kenne:1994:JPCM}. 
But if $J<0.5$ there
is some critical value of $\alpha$ for which $B=\sqrt{2}$ and the 
transition from usual kinks to ${\cal S}$-kinks 
occurs when $\alpha$ decreases (see Fig. \ref{fig:u-n-form}-b). 
And now an interesting question should be raised: what is a
physical {\em single-valued} analogue of the ${\cal S}$-kink in the
{\em discrete} case?

\subsection{Numerical results} 
\label{sec:barrier}

The best remedy to answer the above question 
is to solve Eq. (\ref{sys:eq-un}) numerically. Since for a 
static solution Eq. (\ref{sys:eq-un}) turns itself into a system of
$N$ nonlinear algebraic equations (where $N$ is a number of particles), 
it is convenient to use the Newton-Raphson iterations. To avoid
perturbations due to boundary effects (we use a chain with fixed ends)
the value of $N$ was chosen large enough (typically $N=500$, but it
has been extended to 1000 for broad kinks at small $\alpha$ and big
$J$). The choice of the initial kink form for zeroth iteration is not
very important since for the given problem the Newton-Raphson
iterations are very stable (but to be specific we used Eq. 
(\ref{sol:sol-u-local}) for this purpose). To obtain a stationary 
kink shape with the equilibrium positions $u_n^{\rm eq}$ of the
particles in a chain we usually performed 7--20 iterations. 

\begin{figure}
\centerline{\hbox{
\psfig{figure=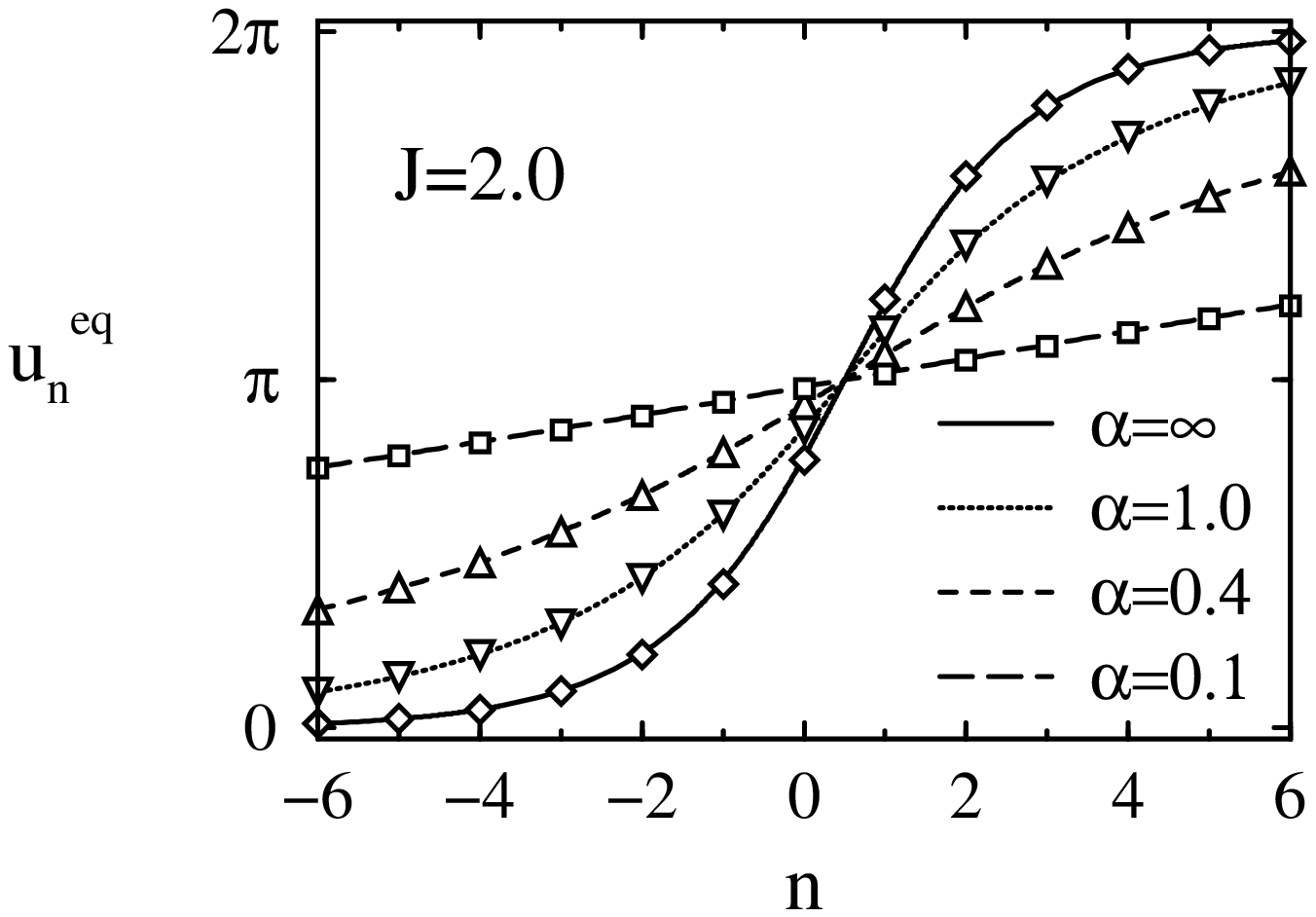,width=80mm,angle=0}}}
\centerline{\hbox{
\psfig{figure=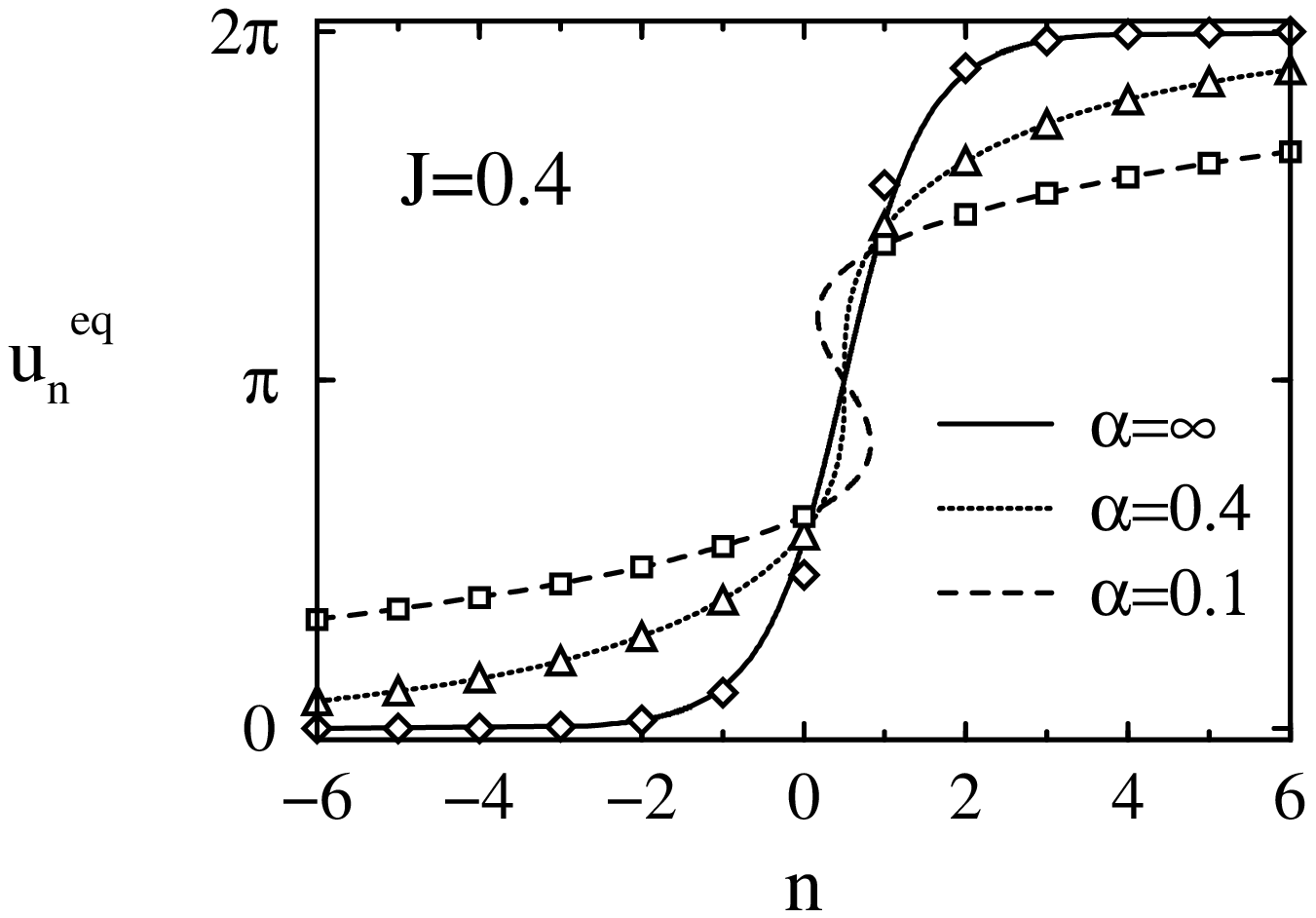,width=80mm,angle=0}}}
\caption{
The comparison of the kink shape predicted analytically with 
that found numerically. Two cases must be distinguished: 
a) $J>0.5$ for which the kink width increases indefinitely 
with decreasing of the $\alpha$; 
b) $J<0.5$ for which the kink width remains finite with 
decreasing of $\alpha$.} 
\label{fig:u-n-form}
\end{figure}

\begin{figure}
\centerline{\hbox{
\psfig{figure=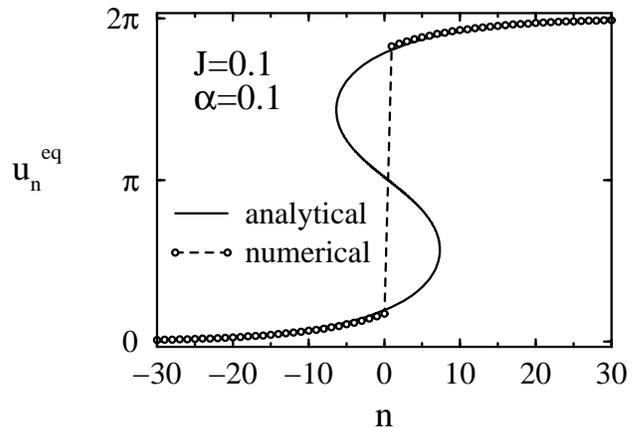,width=80mm,angle=0}}}
\caption{
The kink shape at small $J$ and $\alpha$: instead of the 
predicted analytically multivalued ${\cal S}$-shaped kink 
we obtain numerically the {\em intrinsically localized} kink.} 
\label{fig:u-n-small-J}
\end{figure}

In Figs. \ref{fig:u-n-form}-\ref{fig:u-n-small-J} we compare the
form of the kinks found numerically for different $J$ and $\alpha$ with
the solution (\ref{sol:sol-y})-(\ref{sol:sol-phi}). One can see that
at small $\alpha$ the agreement between them is excellent for all
values of $J$. The only difference is that one should cut out the
unphysical part of the ${\cal S}$-kink and replace it with a vertical
slope to obtain the form of the kink in the discrete case 
(see Fig. \ref{fig:u-n-small-J}). 
In our opinion this result can be understood by reference to
a two-component kink structure. Indeed, as it was shown in Refs. 
\cite{Gaididei:1995:PRL,Gaididei:1997:PD} 
for the anharmonic chain with the Kac-Baker LRI between particles, the
existence of two length scales results into two-component soliton
structure, where the short-range component is dominant in the center
of the soliton, while the long-range component (which can be properly
described in the continuum approximation by Eq. (\ref{sol:eq-B})) 
is dominant in the tails. 

\begin{figure}
\centerline{\hbox{
\psfig{figure=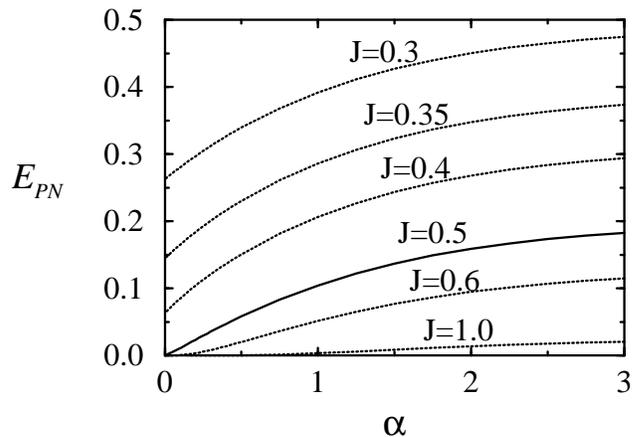,width=80mm,angle=0}}}
\caption{
The dependence of the energy $E_{PN}$ of the Peierls-Nabarro 
barrier on the range $\alpha$ of LRI for different values 
of $J$. The barrier vanishes in the limit $\alpha \to 0$ 
for $J \geq 0.5$ but remains finite for $J<0.5$.} 
\label{fig:peierls}
\end{figure}

Thus, now we can conclude that in the discrete SG model the LRI 
affects the kinks in two opposite ways in relation to the value of
$J$. When $J>0.5$ the increasing of the range of LRI causes the
increasing of the kink width in agreement with the conclusion of Ref. 
\cite{Woafo:1993:JPCM}. But for $J<0.5$ the kinks become {\em
intrinsically localized} as $\alpha \to 0$. In the latter case the form
of the kink is perfectly described by the ${\cal S}$-kink 
in the tails with a vertical slope in the center. 

Numerical calculations show that in both cases the kink energy
monotonically grows to infinity when $\alpha$ decreases to zero. But
the behavior of the Peierls-Nabarro barrier (defined as an energy
difference of the kink centered on a particle and the kink centered
between particles) completely correlates with the behavior of the kink
form (see Fig. \ref{fig:peierls}). When (for $J>0.5$) the kink width 
grows in the limit $\alpha \to 0$, the Peierls-Nabarro barrier
vanishes. But when (for $J<0.5$) the kink becomes intrinsically localized
in this limit, the Peierls-Nabarro barrier remains finite. 

\section{Kink's internal modes} 
\label{sec:internal}

In the previous section we were concentrating upon the static
properties of the kinks. But a considerable interest is also attracted
to the phonon spectrum affected by the presence of the kink. It is
common knowledge that the influence of the kink is to some extent
similar to that of the impurity affecting the phonon spectrum in 
a solid. Namely, not only quantitative changes in the spectrum, but
qualitative ones consisting in emerging the {\em localized modes} 
with frequencies lying beyond the phonon band are to be expected and
present special interest 
\cite{Braun:1997:PRE,Kivshar:1998:PRL,Rice:1983:PRB,Boesch:1990:PRB,Gvozdikova:1998:LTP,Pelinovsky:1998:PD}. 

There should be distinguished localized {\em internal modes} from the
localized {\em translational mode}. The low-frequency translational
mode (which is in the continuum limit the Goldstone mode associated
with the translational invariance) is universally present in an
arbitrary KG model. Its frequency is closely associated with the 
Peierls-Nabarro barrier considered above, so we shall not focus much 
attention on this mode in what follows; instead, we shall consider in
detail the internal modes. These latter play an important role in the
kink dynamics because they can temporarily store
energy taken away from the kink's kinetic energy, which can later be
restored again in the kinetic energy. This gives rise to {\em resonant
structures} in kink interactions 
\cite{Campbell:1983:PD,Peyrard:1983:PD,Campbell:1986:PD,Kivshar:1991:PRL}. 

The most extensively studied is the Rice internal mode
\cite{Rice:1983:PRB} which can be visualized as an oscillation of the
kink's core width. Although this mode does not exist
\cite{Boesch:1990:PRB} in the usual continuum SG model (which is
integrable), even small perturbation of the model brings it into 
the existence 
\cite{Braun:1997:PRE,Kivshar:1998:PRL}. In particular, the Rice mode
exists in the {\em discrete} SG model with NNI \cite{Braun:1997:PRE}.
But the discreteness just changes the dispersion of the system and the
{\em dispersive} LRI under consideration affects it even greater.
Thus, one might expect that the LRI will enhance the creation of the
Rice internal mode in the system being considered. In the next subsection 
we recourse to a variational approach and show that this is so indeed.
Then we investigate kink's internal modes in more detail numerically. 
It turns out that at small $\alpha$ there can exist either {\em
several} kink's internal modes below the phonon spectrum or pronounced
{\em quasilocalized modes} inside the phonon spectrum.

\subsection{Variational approach} 
\label{sec:var}

When, as with the $\varphi^4$--model, the Rice internal mode 
is pronounced it can be properly described by a variational 
collective coordinates approach \cite{Rice:1983:PRB}. 
Proceeding from Eq. (\ref{sys:exp-sum}) and the identity 
\begin{equation}
\sum_n \sum_{m>n} J_{m,n} (u_m-u_n)^2 \equiv 
- \sum_n u_n \sum_{m \neq n} J_{m,n} (u_m-u_n) 
\end{equation}
one can pass to the continuum limit
(\ref{sys:cont}) and write the Hamiltonian (\ref{sys:hamil}) in the
form 
\begin{eqnarray}
\label{var:hamil-x}
H &=& \frac{1}{2 \sinh(\alpha /2)} \int_{-\infty}^{\infty} dy
\biggl\{ \frac{1}{2} \biggl( \frac{du}{dt} \biggr)^2 + (1-\cos u) 
\nonumber \\ 
&-& \frac{1}{2} u(y,t) \frac{J(e^{\alpha}+1) \partial_y^2}{1-\partial_y^2}
u(y,t) \biggr\} \; , 
\end{eqnarray}
where $y= 2 \sinh (\alpha /2) \, n$. 

It should be emphasized that the Rice's collective
coordinates approach \cite{Rice:1983:PRB} cannot be used in our
case. Indeed, choosing the trial function of the form $\phi(y/L(t); B)$,
where $\phi(y;B)$ is the stationary kink solution 
(\ref{sol:sol-y})-(\ref{sol:sol-phi}) and $L(t)$ is a time-dependent
variational parameter (so-called ``effective kink width''), we are unable to
integrate analytically the long-range part of the Hamiltonian 
(\ref{var:hamil-x}). 

Thus, we are forced to introduce another trial function. We call your
attention to the fact (see Fig. \ref{fig:phi-y-form}) that the change
of $B$ in the kink solution $\phi(y;B)$ changes a slope of the kink and 
its width as well. Therefore, to describe small-amplitude kink
oscillations around its stationary form one can use equally well 
instead of Rice's trial function a trial function of the form
\begin{equation}
\label{var:trial}
u_n(t) = \phi ( 2 \sinh(\alpha /2) \, n \, ; \, b(t) ) \; , 
\end{equation}
where $\phi$ is determined by Eqs.
(\ref{sol:sol-y})-(\ref{sol:sol-phi}) and $b(t)$ is the 
time-dependent variational parameter. Then, using that 
$\phi(y;b(t))$ is the solution of the equation 
\begin{equation}
\label{var:eq-bt}
\frac{\partial_y^2}{1-\partial_y^2} \phi(y;b(t)) = 
\frac{\sin \phi(y;b)}{b^2-1} \; , 
\end{equation}
the integrals appearing in Eq. (\ref{var:hamil-x}) can be taken
analytically: 
\begin{equation}
\label{var:int-u}
U(b)=\int_{-\infty}^{\infty}dy (1-\cos\phi)=
4b-2\log \biggl( \frac{b+1}{b-1} \biggr) \, ,
\end{equation}
\begin{equation}
\label{var:int-v}
V(b)=-\frac{1}{2}\int_{-\infty}^{\infty}dy \phi \sin \phi
=2b+(b^2-1)\log \biggl( \frac{b+1}{b-1} \biggr) \, , 
\end{equation}
and
\begin{eqnarray}
\label{var:int-m}
M(b) &=& \int_{-\infty}^{\infty}dy \biggl( \frac{d\phi}{db}
\biggl)^2 = \frac{2}{b^2-1} 
\nonumber \\ &\times&
\int_0^{1/b}d\mu \frac{ \biggl[ 2b\mu+(b^2-1)\log \biggl( 
\frac{1+b\mu}{1-b\mu}
\biggr) \biggr]^2}{(1-\mu^2) \, (b^2-2+b^2\mu^2)} \, ,
\end{eqnarray}
where the last integral was taken analytically as well, yet its
cumbersome expression prevents us from writing it down here in an
explicit form. We just mention that $M(b)$ grows to infinity as
$b\to\sqrt{2}$ and has the asymptotics
\begin{equation}
\label{var:int-m-assym}
M(b)\simeq \frac{2\pi^2}{3b} + (12 + \pi^2) \frac{2}{3b^3} 
+ (16+\frac{46}{45} \pi^2) \frac{1}{b^5} 
+ {\cal O}\biggl(\frac{1}{b^7} \biggr) 
\end{equation}
for big values of $b$. 

Thus the effective Hamiltonian of the system takes on the form 
\begin{eqnarray}
\label{var:hamil-eff}
H_{eff} = \frac{1}{2 \sinh(\alpha /2)} \biggl\{ \frac{1}{2} M(b) 
\biggl( \frac{d b}{dt} \biggr)^2 + W(b) \biggr\} \; , 
\end{eqnarray}
where the potential energy of the kink 
\begin{eqnarray}
\label{var:en-series}
W(b) &=& U(b)+ \frac{B^2-1}{b^2-1} \, 
V(b) \simeq  E_{K}(B) \nonumber \\ 
&+& \frac{4B^3}{(B^2-1)^2}(b-B)^2 + {\cal O}(b-B)^3 \, ,
\end{eqnarray}
as it would be expected, is minimal in the point 
$b=B \equiv \sqrt{1+J(e^{\alpha}+1)}$, where it equals to the energy 
\begin{eqnarray}
\label{var:en-kink}
E_{K}(B) = 6B+(B^2-3)\log \biggl( \frac{B+1}{B-1} \biggr) 
\end{eqnarray}
of the stationary kink. The kink energy (\ref{var:en-kink}) was
calculated first (although in a more bulky form) in
Ref. \cite{Woafo:1993:JPCM}. 

\begin{figure}
\centerline{\hbox{
\psfig{figure=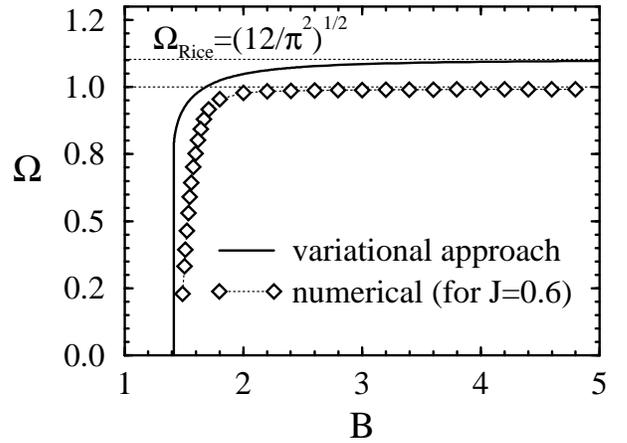,width=80mm,angle=0}}}
\caption{
The dependence of the frequency $\Omega$ of the Rice's 
kink internal mode on the parameter of the nonlocality 
$B=\sqrt{1+J(e^{\alpha}+1)}$: predicted analytically 
by Eq. \protect\ref{var:freq} (full line) and found numerically 
for $J=0.6$ (diamonds).}
\label{fig:freq}
\end{figure}

For small deviations of $b$ from $B$ (in the harmonic
approximation) the effective Hamiltonian (\ref{var:hamil-eff})
generates the equation of motion 
\begin{equation}
\label{var:eq-osc}
\biggl( \frac{d^2}{dt^2}+\Omega^2 \biggr) (b(t)-B)=0
\end{equation}
being the equation of motion for the harmonic oscillator 
with the frequency
\begin{equation}
\label{var:freq}
\Omega=\biggl[ \frac{8B^3}{M(B)(B^2-1)^2} \biggr]^{1/2} \, ,
\end{equation}
whose dependence on $B$ is depicted on Fig. \ref{fig:freq}. 

Thus, a slightly excited kink will oscillate around its stationary
shape with the frequency $\Omega$, depending upon the parameter 
$B$ which specifies the nonlocality of the system. 
When $B$ decreases (nonlocality grows) the frequency $\Omega$
also decreases ($\Omega \to 0$ when $B \to \sqrt{2}$). Although Eq.
(\ref{var:freq}) is not fully precise even in the continuum limit (to
improve it an interaction with phonons should be taken into account
\cite{Boesch:1990:PRB}) it is in a good agreement with the numerical
calculations (see Fig. \ref{fig:freq}) which are described below.

\subsection{Numerical results} 
\label{sec:modes}

\begin{figure}
\centerline{\hbox{
\psfig{figure=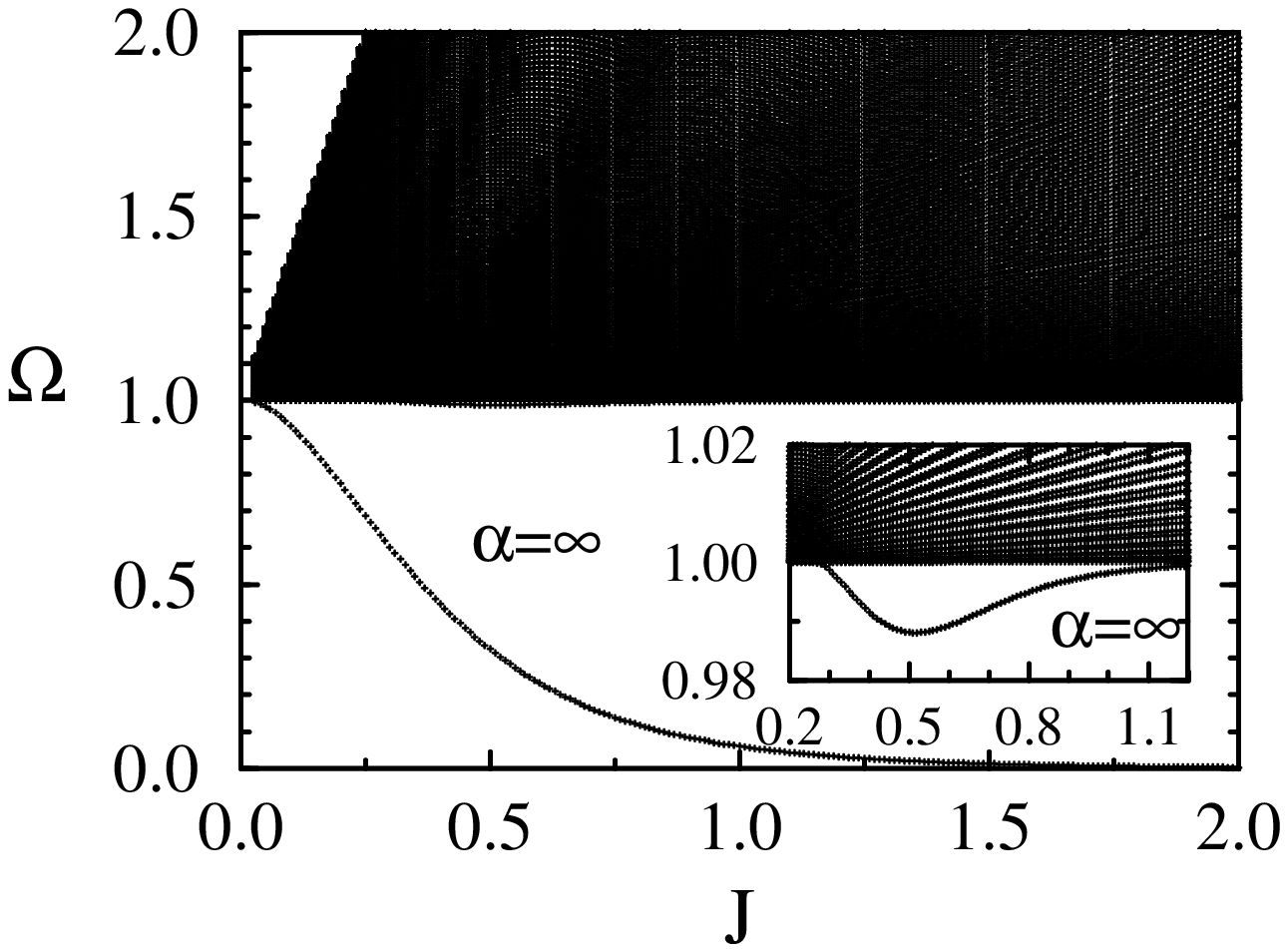,width=80mm,angle=0}}}
\vspace{1mm}
\centerline{\hbox{
\psfig{figure=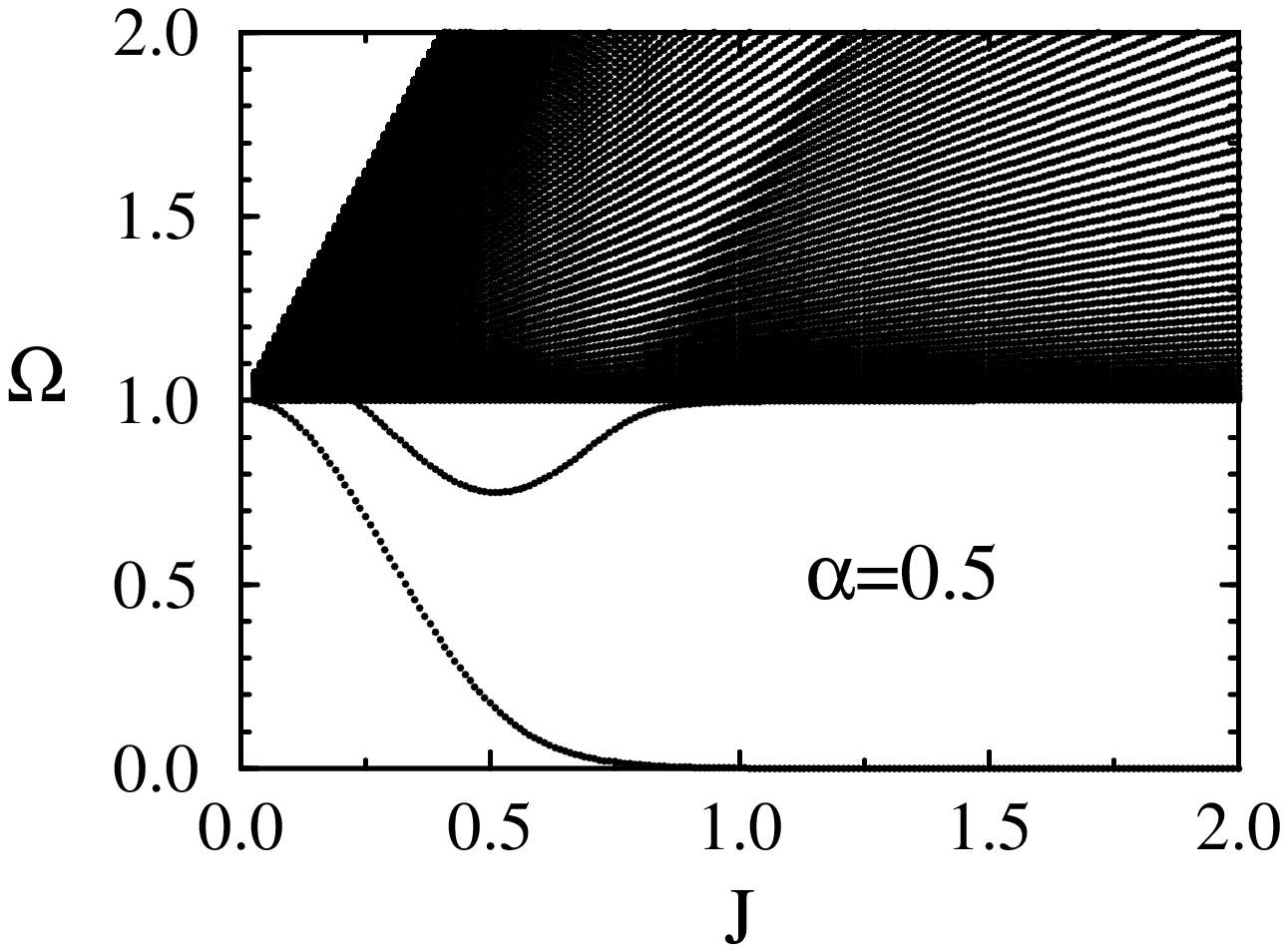,width=80mm,angle=0}}}
\vspace{5mm}
\caption{
Spectrum of small-amplitude excitations around a kink as a function of
the coupling parameter $J$ for $\alpha=\infty$ and $\alpha=0.5$.}
\label{fig:spectrum-big-alpha}
\end{figure}

\begin{figure}
\centerline{\hbox{
\psfig{figure=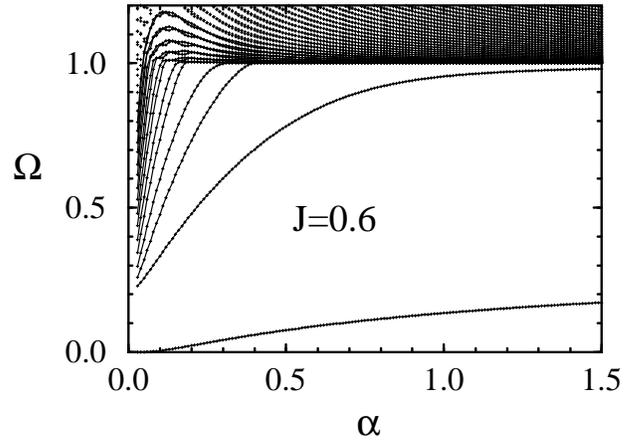,width=80mm,angle=0}}}
\vspace{7mm}
\caption{Spectrum of small-amplitude excitations around a kink as a function of
the nonlocality parameter $\alpha$ for $J=0.6$.}
\label{fig:spectrum-J06}
\end{figure}

\begin{figure}
\centerline{\hbox{
\psfig{figure=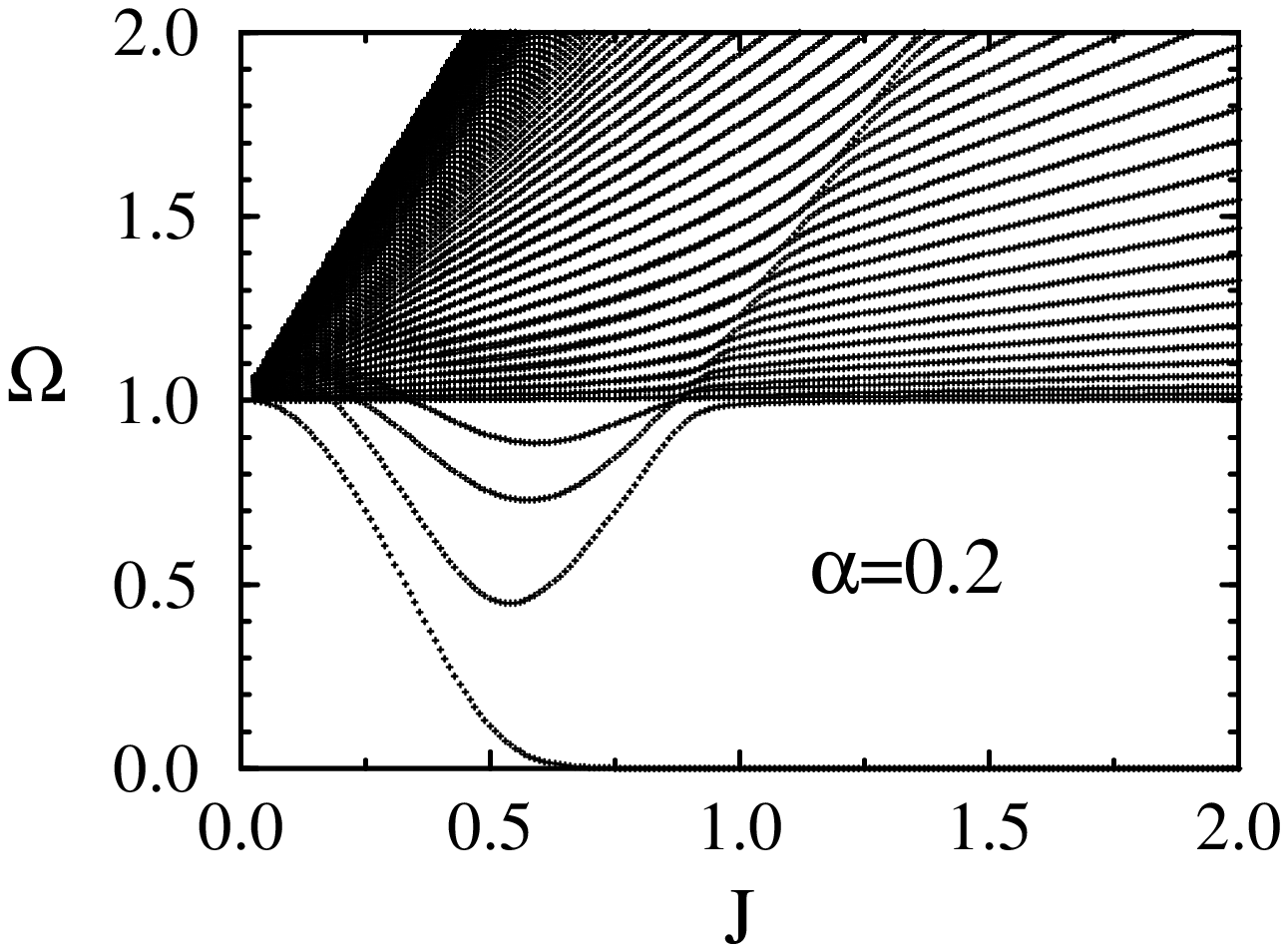,width=80mm,angle=0}}}

\centerline{\hbox{
\psfig{figure=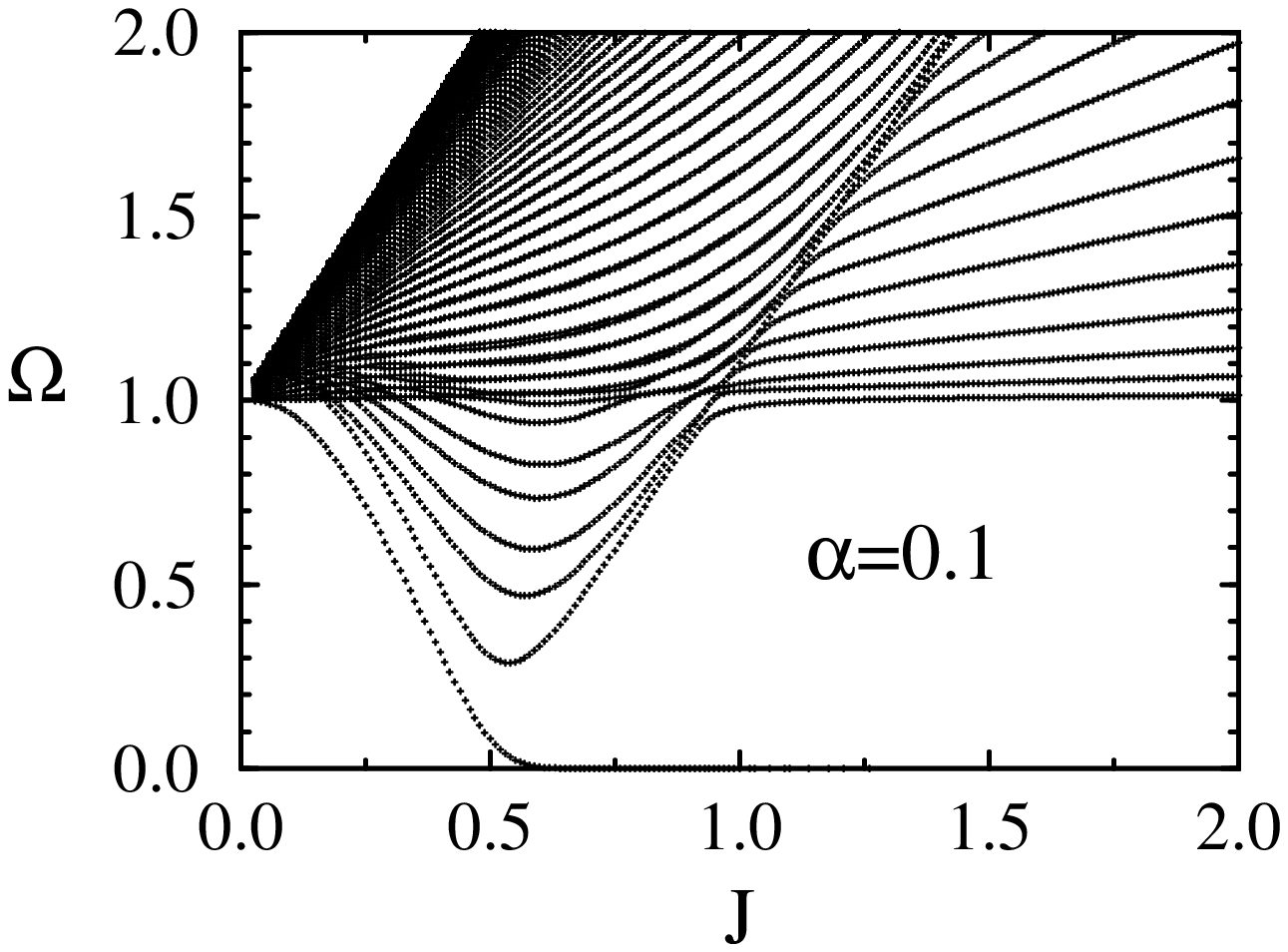,width=80mm,angle=0}}}
\caption{Spectrum of small-amplitude excitations around a kink 
as a function of the coupling parameter $J$ for: 
a) $\alpha=0.2$ and b) $\alpha=0.1$.}
\label{fig:spectrum-small-alpha}
\end{figure}

\begin{figure}
\centerline{\hbox{
\psfig{figure=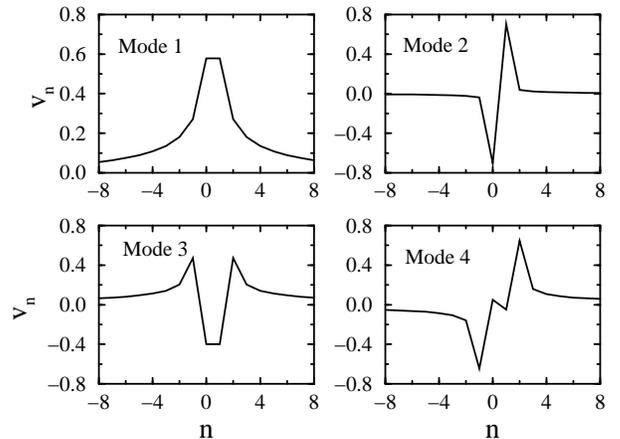,width=80mm,angle=0}}}
\caption{
Example of the shape of the lowest localized modes: the 
translational mode (Mode 1), the Rice mode (Mode 2), and so on.}
\label{fig:local-modes}
\end{figure}

\begin{figure}
\centerline{\hbox{
\psfig{figure=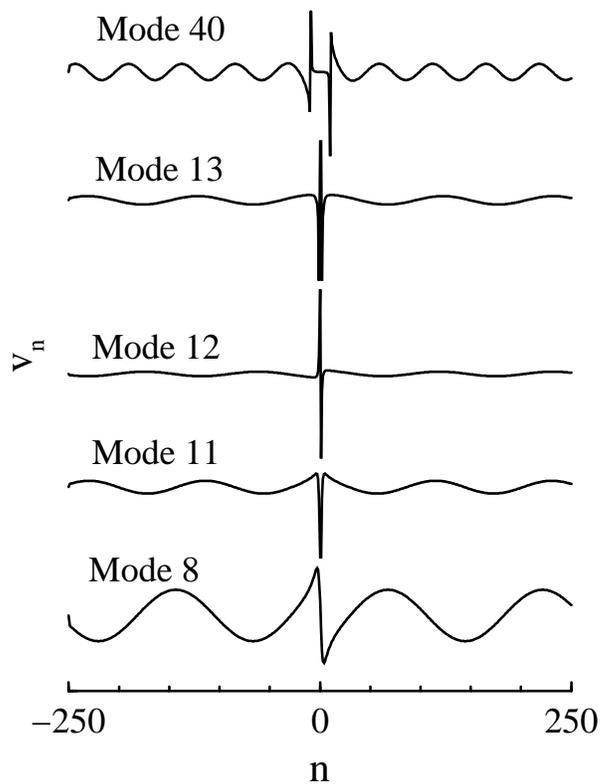,width=80mm,angle=0}}}
\caption{The shape of the delocalized and quasilocalized states 
inside the phonon band for $J=1.25$ and $\alpha=0.1$.}
\label{fig:quasilocal-modes}
\end{figure}

\begin{figure}
\centerline{\hbox{
\psfig{figure=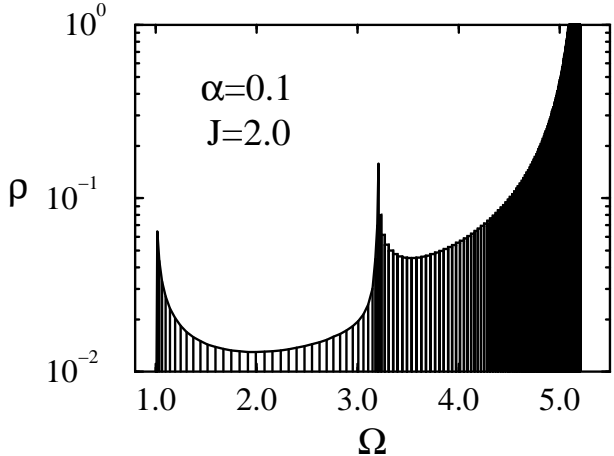,width=80mm,angle=0}}}
\caption{The density of the eigenstates in the phonon band as 
a funcion of the eigenfrequency for $J=2.0$ and $\alpha=0.1$.}
\label{fig:spectral-density}
\end{figure}

The exposed variational approach (even in a form complicated by
introducing of several time-dependent parameters) permits one to
investigate only a limited number of oscillatory modes. To overview
rather the entity of the whole phonon spectrum one should deal with
the initial set of the equations of motion (\ref{sys:eq-un}).
Specifically, when all the equilibrium positions $u_n^{\rm eq}$ 
of the particles in a chain with a kink become known by the method
described in Sec. \ref{sec:barrier}, we can study the spectrum of
small-amplitude oscillations around this state by looking for a
solution of Eq. (\ref{sys:eq-un}) in the form
\begin{equation}
\label{mod:vn}
u_n = u_n^{\rm eq} + v_n \, e^{i\Omega t} \; . 
\end{equation}
We assume that the deviations $v_n$ of the particles from the kink
shape are sufficiently small ($v_n \ll a$) and ignore all nonlinear
terms in the equations of motion for $v_n$. Introducing this ansatz
into Eq. (\ref{sys:eq-un}), we obtain a system of linear equations
that can be written in a matrix form 
\begin{equation}
\label{mod:eq-vn}
\hat{D} \vec{v} = \Omega^2 \vec{v} \; , 
\end{equation}
where $\vec{v} \equiv \{v_n\}$ and a symmetric matrix $\hat{D}$ is
the dynamical matrix of the lattice in the presence of a kink with the
components: 
\begin{eqnarray}
\label{mod:Dnm}
D_{n,n} &&= 2J + \cos u_n^{\rm eq} \; , \nonumber \\ 
D_{n,m} &&= D_{m,n} = - J_{n,m} \; . 
\end{eqnarray}
By virtue of the fact that the matrix $\hat{D}$ is symmetric all its
eigenvalues are real. They give us the frequencies of the
small-amplitude oscillations around the kink while the corresponding
wavevectors describe the spatial profile of each mode. The eigenvalues
and eigenvectors of the matrix $\hat{D}$ were calculated numerically
using the Householder matrix reduction to tridiagonal form and the
$QL$ diagonalization algorithm \cite{Press:1997:NRC}. 
The obtained results are presented in Figs. 
\ref{fig:spectrum-big-alpha} - \ref{fig:spectral-density}. 
In the following discussion of these results, we assume that the
reader is familiar with Ref. \cite{Braun:1997:PRE} where the basic
features of the kink's linear spectrum are comprehensively expounded. 

\vspace{5mm}

In Fig. \ref{fig:spectrum-big-alpha} we compare the kink's linear
spectrum $\Omega(J)$ for $\alpha=0.5$ with the spectrum obtained in
Ref. \cite{Braun:1997:PRE} for the NNI limit ($\alpha=\infty$). As it 
has there been shown, in the discrete SG model with
the interaction between nearest neighbors aside from the low-frequency
translational localized mode there exist (slightly expressed in the
interval $0.27 \lesssim J \lesssim 1.59$) the Rice internal mode. 
One can see from Fig. \ref{fig:spectrum-big-alpha} that for $\alpha=0.5$
the situation remains qualitatively the same, except that the Rice
mode becomes evidently pronounced. Thus, in agreement with the results
of the variational approach, the LRI strongly enhances creation of 
the Rice mode (see also Fig. \ref{fig:freq}). 

\vspace{15mm}

Furthermore, when $\alpha$ yet decreases, there occur new qualitative 
phenomena. Foremost, additional localized internal modes are split out
from the bottom of the phonon spectrum. This process is easily
observable in Fig. \ref{fig:spectrum-J06} where the kink's linear
spectrum $\Omega$ as a function of $\alpha$ is plotted for $J=0.6$.
One can see that the number of localized modes grows indefinitely as
$\alpha$ vanishes. In particular, there are four localized modes at
$\alpha=0.2$ 
(see Fig. \ref{fig:spectrum-small-alpha}-a) 
and seven at $\alpha=0.1$ 
(see Fig. \ref{fig:spectrum-small-alpha}-b).
All the localized internal modes are best pronounced around $J \simeq
0.6$. Closer analytical examination of Eqs. 
(\ref{mod:eq-vn})-(\ref{mod:Dnm}) shows that all eigenstates are
either symmetric or antisymmetric. One can state on the strength of
the numerical calculations that the symmetric and antisymmetric states
are always alternating, starting with the symmetric translational mode
(Mode 1 in Fig. \ref{fig:local-modes}) and antisymmetric Rice mode 
(Mode 2 in Fig. \ref{fig:local-modes}).

Another interesting feature that appears in the kink's linear spectrum
at small $\alpha$ is that there exist an apparent transformation of
the phonon spectrum (in the plane $\Omega - J$) along an imaginary
line (let us call it ``ridge'') which extends the localized internal
modes inside the phonon spectrum in the direction of larger $J$ 
(see Fig. \ref{fig:spectrum-small-alpha}). This ridge is evidently
parallel to the upper cut-off frequency of the phonon band, with the
spectrum curves looking like essentially different left or right from
the ridge. 
In the domain right to the ridge each symmetric eigenstate is
equidistant from both nearest antisymmetric eigenstates. 
In the domain left to 
the ridge each symmetric eigenstate is paired with an appropriate
antisymmetric eigenstate: their frequencies become almost coincident. The
shapes of the symmetric and antisymmetric states in a pair are also in
a close correlation: $v_n^{\rm a} \simeq \theta(n-n_0) \, v_n^{\rm s} \,$, 
where $n_0$ is the position of the kink center and $\theta(x)$ is the 
Heaviside function. 
The ridge is shifted towards the upper cut-off frequency as
$\alpha$ grows and completely disappears (or rather coincides with the
upper edge of the phonon band) in the NNI limit ($\alpha=\infty$).

The most interesting phenomenon however concerns the shape of the
eigenstates in the vicinity of the ridge (see Fig.
\ref{fig:quasilocal-modes}). One can see that both above 
(e.g., mode 40 in Fig. \ref{fig:quasilocal-modes}) 
and below 
(e.g., mode 8 in Fig. \ref{fig:quasilocal-modes})
the ridge the eigenstates are delocalized 
as it is expected to be in the continuum spectrum. 
But the shape of the eigenstates undergoes essential
changes when approaching the ridge frequency. Namely, such eigenstates
practically have no distortion in the oscillatory subspace and thus
are {\em quasilocalized} 
(see modes 11--13 in Fig. \ref{fig:quasilocal-modes}).
It should be indicated a similarity between these quasilocalized
states and the ``exotic states'' investigated in Refs. 
\cite{Eiermann:1991:JL,Eiermann:1992:JCP,Sonnek:1995:PRB}, although we
could not yet provide an in-depth analysis of this analogy. 

It is notable that the quasilocalization of the eigenstates in the 
vicinity of the ridge frequency is accompanied by pronounced {\em
densening} of the eigenstates which can be viewed in Fig. 
\ref{fig:spectral-density} where we plot (at fixed $\alpha$ and $J$)
the density of the eigenstates as function of frequency. Because of
this peculiarity inside the phonon spectrum, the ridge of
quasilocalized states should be observable experimentally. We expect
that similar to the localized internal modes, the quasilocalized
states could play a part in the kink dynamics.

\section{Radiation of moving kinks} 
\label{sec:radiation}

All the preceding was devoted to the properties of immobile 
kinks. In this section we go out of this restriction and consider
effects of the LRI on the moving kinks. To avoid discreteness effects
we assume that $J>0.5$ and $\xi \gg 1$, so that the kinks are broad
and the Peierls-Nabarro barrier can be neglected. In other words we
assume that the kinks are properly described by Eq. (\ref{sys:eq-uz}).
It is well known that in the limit of short-ranged dispersion 
($\sigma=0$ in Eq. (\ref{sys:eq-uz})) the kinks can move without
changing their shape and velocity. In fact, it is a consequence of the
Lorentz invariance of Eq. (\ref{sys:eq-uz}) for $\sigma=0$. But this
invariance is broken by the LRI (as well as it would be broken by the
discreteness). Therefore, one might expect that similar to the
discreteness \cite{Combs:1983:PRB} the LRI causes a radiation of 
moving kinks.
Indeed, this effect was recently shown to exist for fluxons in a
superlattice of Josephson junctions \cite{Gaididei:SSJJ} which is
described by another variant of nonlocal SG equation. In what follows
we extend the results of Ref. \cite{Gaididei:SSJJ} to our model. 

To investigate propagating solutions we introduce 
the transformation to the moving frame of reference 
in which the center of the kink is at rest:
\begin{equation}
\label{rad:defs}
\zeta = (z-vt) \, \gamma(v) \quad , \quad 
\tau=(t-vz) \, \gamma(v) \; , 
\end{equation}
where $\gamma(v)=1/\sqrt{1-v^2}$ is the Lorentz 
factor and $v$ is the dimensionless kink velocity. 
Applying this transformation to Eq. (\ref{sys:eq-uz}) 
we obtain
\begin{eqnarray}
\gamma^2 (\partial^2_{\tau} && - 2v\partial_{\tau} 
\partial_{\zeta} + v^2 \partial^2_{\zeta}) \, u(\zeta, \tau) 
\nonumber \\ 
&& + \frac{\gamma^2 \hat{k}^2(v)}{1+\sigma^2 \gamma^2 \hat{k}^2(v)} 
\, u + \sin u = 0 \; ,
\label{rad:eq-theta-tau}
\end{eqnarray}
where the notation 
\begin{equation}
\hat{k}^2(v) = 2v \partial_{\tau} \partial_{\zeta} - 
\partial^2_{\zeta} - v^2 \partial^2_{\tau} 
\end{equation}
was used. We introduce 
\begin{equation}
\label{rad:u-phi-f}
u(\zeta,\tau)=\phi(\zeta/\sigma\gamma; B) + f(\zeta,\tau) \; ,
\end{equation}
where $\phi(\zeta/\sigma\gamma; B)$ is the  static solution 
given by Eqs. (\ref{sol:sol-y})-(\ref{sol:sol-phi}) and the
function $f(\zeta,\tau)$ describes the change of the kink 
shape and the radiation. Inserting Eq. (\ref{rad:u-phi-f}) 
into Eq. (\ref{rad:eq-theta-tau}) in the linear approximation 
with respect to $f(\zeta,\tau)$ we obtain for the Fourier 
transform of $f(\zeta,\tau)$
\begin{equation}
\bar{f}(k,\tau)=\frac{1}{2\pi} \int_{-\infty}^{\infty} 
e^{-i\,k\,\zeta} \, f(\zeta,\tau) \, d \zeta 
\end{equation}
the inhomogeneous differential equation 
\begin{eqnarray}
&& \gamma^2 (\partial^2_\tau - 2i\,v\,k \partial_\tau) \, 
\bar{f} + \Omega^2(k,v) \, \bar{f} 
\nonumber \\
&& + \left( \frac{\gamma^2 \hat{k}^2(v)}{1+\gamma^2 \sigma^2 
\hat{k}^2(v)} - \frac{\gamma^2 k^2}{1+\gamma^2 \sigma^2 k^2} 
\right) \, \bar{f} 
\nonumber \\
&& - \overline{(1-\cos\phi)\,f} = - i \, k \gamma^2 v^2 \,\, 
\overline{(\partial_\zeta\,\phi)}(k) \; , 
\label{rad:eq-k-tau}
\end{eqnarray}
where
\begin{equation}
\hat{k}^2(v)= k^2+ 2 i \, vk \, \partial_\tau - 
v^2 \partial^2_\tau 
\end{equation}
and the function 
\begin{eqnarray}
\Omega^2(k,v)= 1 - \gamma^2 v^2 k^2 + 
\frac{\gamma^2 k^2}{1+\gamma^2 \sigma^2 k^2} 
\end{eqnarray}
is the dispersion relation in the moving frame of reference. 
The right-hand-side of Eq. (\ref{rad:eq-k-tau}) is the source 
of the radiation. It vanishes when the kink velocity  $v=0$.

The equation $\Omega(k,v)=0$ has two real roots $k=\pm\,k_r$. 
The existence of these roots means 
\cite{Kuehl:1990:PFB,Karpman:1993:PRE,Gaididei:1996:PLA} 
that waves with wavenumbers $\pm\,k_r$ will be resonantly 
excited by a kink, forming an oscillatory tail. 
We restrict ourselves to the investigation of small kink 
velocities for which the resonant wavenumber 
$k_r$ can be expressed as follows 
\begin{eqnarray}
k_r \simeq \frac{B}{v} \; . 
\end{eqnarray}
In the resonant region $k \simeq\,\pm\,k_r$ we can neglect 
the second derivative $\partial^2_\tau \,f$
because as it will be seen later the function $f$ slowly changes 
with $\tau$. We use the expansion 
\begin{eqnarray}
\Omega^2(k,v) \simeq 2 \gamma^2 v^2 k_r \, (k_r \mp k) 
\quad \mbox{for} \quad k \simeq \pm k_r
\label{rad:Omega-approx}
\end{eqnarray}
and replace the right-hand side of Eq. (\ref{rad:eq-k-tau}) by 
its value at $k= \pm k_r$. We represent also the function 
$\bar{f}(k,\tau)$ as a sum
\begin{equation}
\bar{f}=\bar{f}_{+}+\bar{f}_{-} \; , 
\label{rad:fp-fm}
\end{equation}
where the function $\bar{f}_+ \,$ ($\bar{f}_-$) differs from 
zero only in the close vicinity of the wave vector $k_r \,$ 
($-k_r$). Taking into account Eqs. 
(\ref{rad:Omega-approx})-(\ref{rad:fp-fm}) we 
obtain from Eq. (\ref{rad:eq-k-tau}) that the functions 
$\bar{f}_{\pm}$ satisfy the equation
\begin{eqnarray}
i \, \partial_\tau \, \bar{f}_{\pm} &+& v (k \mp k_r) 
\bar{f}_{\pm} \nonumber \\ 
&\pm& \frac{1}{2B} \, \overline{(1-\cos\phi) \, f_{\pm}} 
= \frac{i}{2} \, v \, A \; , 
\label{rad:eq-kr-tau}
\end{eqnarray}
where 
\begin{equation}
A= \overline{(\partial_{\zeta} \phi)} (k_r) \approx 
\sech \left( \frac{\pi B}{2 \, v} \right) \; . 
\label{rad:A}
\end{equation}
Returning to the real space $(\zeta,\tau)$ we obtain from 
Eq. (\ref{rad:eq-kr-tau}) 
\begin{eqnarray}
(\partial_\tau - v \partial_\zeta) f_{\pm} 
\pm i \left[ B - \frac{1}{2B} (1-\cos \phi) \right] f_{\pm} 
= \frac{v A}{2} \, \delta(\zeta) \; .
\label{rad:eq-theta-tau-lin}
\end{eqnarray}
With the initial condition $f(\zeta,\tau)/_{\tau=0}=0$, 
the solution of Eq. (\ref{rad:eq-theta-tau-lin}) 
becomes
\begin{equation}
f=A \left(\theta(\zeta + v \tau) - \theta(\zeta) \right) 
\cos \left(k_r \, \zeta - \chi(\zeta) \right) \; , 
\end{equation}
where 
\begin{eqnarray}
\chi(\zeta) &&= \frac{1}{2Bv} \int_{0}^{\zeta} 
\left[ 1-\cos \phi(\zeta'/\sigma \gamma ; B) \right] 
\, d \zeta' \nonumber \\ 
&& \approx \frac{1}{B \, v} \tanh(\zeta) 
\end{eqnarray}
is the phase function. 
It can be seen that $|k_r\zeta|\gg\,\chi(\zeta)$ when $|\zeta|>1$. 
Therefore, we can neglect the phase function $\chi(\zeta)$ and 
returning to the original variables $(z,t)$ we obtain that the 
radiation of the kink which moves with a small velocity $v$ is 
given by the function
\begin{equation}
f(z,t)= A \left[ \theta(z) - \theta(z-vt) \right] \, 
\cos[(B/v)\, z - B \, t] \; , 
\end{equation}
with the amplitude of the radiation $A$ which exponentially decreases
when the kink velocity tends to zero. 

Thus, now one can conclude that the Kac-Baker LRI produces 
{\em non-perturbative} radiation with the wave length 
$\lambda=2\pi/k_r \sim v$ that emerges in the rear of the kink.

\section{Conclusions} 
\label{sec:summary}

We have investigated the effects of the Kac-Baker long-range
interaction on the kink's properties in the discrete sine-Gordon
model. We have obtained an implicit form for the kink's shape and
energy and have shown that the kink width increases 
indefinitely as $\alpha$ vanishes only 
in the case of strong interparticle coupling ($J>J_{cr}=0.5$). 
On the contrary, the kink becomes {\em intrinsically 
localized} for $J(e^{\alpha}+1)<1$. 
Accordingly, we have shown that the Peierls-Nabarro barrier 
vanishes as $\alpha \to 0$ for supercritical values of 
the coupling $J$ but remains {\em finite} for subcritical values. 
We have developed a new variant of the collective coordinates 
variational approach for investigation of the internal kink's 
dynamics and have shown that the Kac-Baker LRI essentially 
enhances creation of the kink's {\em internal localized modes}. 
We have demonstrated numerically that indefinite number of the
localized internal modes came into existence when $\alpha$ 
approaches zero. We have revealed an existence of the 
{\em quasilocalized states} inside the phonon spectrum 
for small $\alpha$ and large $J$. We have described their
properties, in particular, a pronounced densening of the 
phonon spectrum in the vicinity of the quasilocalized states. 
We expect that similar to the localized internal modes, the 
quasilocalized states could play a part in the kink dynamics. 
We have investigated moving kinks in the continuum limit 
and have shown that, due to break of the Lorentz invariance by 
the LRI, they always {\em radiate plane waves} with the wave 
length proportional to the kink velocity.

\section*{Acknowledgments}

S.M. and S.Sh. thank the Department of Theoretical 
Physics of the Palack\'y University in Olomouc for the 
hospitality. 
S.M., E.M. and S.Sh. acknowledge support from the 
Grant Agency of Czech Republic (Grant No.~202/98/0166). 
S.M. and Yu.G. acknowledge partial support from 
the DLR project UKR--002--99.



\end{multicols}
\end{document}